\documentclass[superscriptaddress,nofootinbib,notitlepage]{revtex4}
\usepackage{amsfonts}
\usepackage[bookmarksopen,colorlinks]{hyperref}
\usepackage{graphicx}
\usepackage{dcolumn}
\usepackage{bm}
\usepackage{amssymb,amsmath}
\usepackage[dvips]{color}

\begin{document}

\title{{\Large Kerr geometry in $f(T)$ gravity }}
\author{Cecilia Bejarano}
\email[]{cbejarano@iafe.uba.ar} \affiliation{Instituto de
Astronom\'ia y F\'isica del Espacio (IAFE, CONICET-UBA), Casilla de
Correo 67, Sucursal 28, 1428 Buenos Aires, Argentina.}
\author{Rafael Ferraro}
\email[Member of Carrera del Investigador Cient\'{\i}fico (CONICET,
Argentina); ]{ferraro@iafe.uba.ar} \affiliation{Instituto de
Astronom\'ia y F\'isica del Espacio (IAFE, CONICET-UBA), Casilla de
Correo 67, Sucursal 28, 1428 Buenos Aires, Argentina.}
\affiliation{Departamento de F\'isica, Facultad de Ciencias Exactas
y Naturales, Universidad de Buenos Aires, Argentina.}
\author{Mar\'ia Jos\'e Guzm\'an}
\email[]{mjguzman@iafe.uba.ar} \affiliation{Instituto de
Astronom\'ia y F\'isica del Espacio (IAFE, CONICET-UBA), Casilla de
Correo 67, Sucursal 28, 1428 Buenos Aires, Argentina.}

\begin{abstract}
Null tetrads are shown to be a valuable tool in teleparallel theories of
modified gravity. We use them to prove that Kerr geometry remains a solution
for a wide family of $f(T)$ theories of gravity.
\end{abstract}

\maketitle


\section{Introduction}

\label{sec:intro}

Many models of modified gravity have been proposed in order to tackle the
shortcomings of General Relativity (GR). Deformations of the Einstein's
theory at small or large scales, i.e. ultraviolet or infrared gravity, could
provide a better handling of singularities and the cosmic acceleration. In
particular, a proper deformation of GR in the ultraviolet regime could play
the role of describing the transition between GR and quantum gravity. This
frontier could be better understood by resorting to a \textit{teleparallel}
formulation of gravity. As a matter of fact, although with a different
purpose in mind, it was Einstein himself who proposed in the 30's the
reformulation of GR in a teleparallel framework, by taking the field of
orthonormal frames or \emph{tetrads} as the dynamical variable instead of
the metric tensor \cite{Ein}. Certainly, teleparallelism is part of a bigger
picture. Gravity can be described by means of a connection having both
curvature (like GR, through the Levi-Civita torsionless connection) and
torsion (like teleparallelism, through the Weitzenb\"{o}ck curvatureless
connection) leading to Einstein-Cartan theory \cite{Car22, Heh76}.

In this article we will focus on the so-called $f(T)$ gravity, a theory of
modified gravity based on a spacetime possessing absolute parallelism \cite%
{Fer07, Fer08,Ben09,Bam10, Lin10, Wu10, Ben11,Li11a, Fer11a, Fer11b, Sot11,
Wei11, Zhe11}. The first results of this alternative theory of gravity were
very encouraging. It has been shown that the teleparallelism \`{a} la
Born-Infeld cures the primordial singularity of flat Friedman-Lema\^{\i}%
tre-Robertson-Walker (FLRW) universes, providing a natural inflationary
early stage without invoking a new field \cite{Fer07, Fer08}. An extension
of $f(T) $ gravity \cite{Fio14} changed the conical singularity of a 3D
cosmic string into a geodesically complete smooth curved spacetime \cite%
{Fer10}. Like other theories of modified gravity, $f(T)$ theories display
additional degrees of freedom; in this case they come from the loss of local
Lorentz invariance and their physical nature is not well understood yet \cite%
{Li11a, Sot11,Yan10,Li11b,Li11}.

A remarkable feature of $f(T)$ theories is that the dynamics of tetrads is
described by second order equations, which is not usual in the context of
modified gravity. This property is guaranteed by the teleparallel
Lagrangian, which is a function of the square of the first derivatives of
the tetrad field (differing from GR, whose Lagrangian contains second
derivatives of the metric). The teleparallel Lagrangian is built in a
Weitzenb\"{o}ck spacetime, i.e., a spacetime endowed with a curvatureless
connection proportional to first derivatives of the tetrad. The central
piece of a teleparallel Lagrangian is the Weitzenb\"{o}ck torsion. The
simplest teleparallel theory is the teleparallel equivalent of GR (TEGR),
which is just Einstein's gravity in the language of tetrads. $f(T)$ theories
are defined by deformations of the TEGR Lagrangian.

A main point to consider in modified gravity is the possibility of smoothing
black hole singularities. The searching for solutions of spacetimes
displaying spherical or axial symmetry is not trivial in $f(T)$ gravity. In
fact, usually one uses the symmetry for choosing coordinates such that the
metric looks simple. Even so, there are many tetrads for a given metric.
Since $f(T)$ theories are not invariant under local Lorentz transformations,
the knowledge of the metric symmetry gives no idea of the ansatz for the
tetrad field. In this article we will study vacuum axially symmetric
rotating solutions of $f(T)$ gravity. We are going to prove that Kerr
geometry \cite{Ker63} remains a solution for $f(T)$ gravity. We will employ
null tetrads as a useful tool for straightforwardly getting the result. We
remark that different aspects of rotating black holes have been discussed in
the context of TEGR \cite{Per01, Dar03,Mal02,Mal07,Tia09}. Among the papers
about Kerr spacetime in modified gravity, we can mention those referred to $%
f(R)$ gravity \cite{Cap10,Myu11,Myu13,Per13}. Other rotating geometries,
like the G\"{o}del universe \cite{God49}, have been also studied in TEGR
\cite{Obu04, Sou10}, as well as in $f(R)$ theories of gravity in both metric
and Palatini approaches \cite{Cli05,Reb09,San10}. Recently, the rotating
cosmology was also explored in $f(T)$ gravity \cite{Liu12}.

The paper is organized as follows. In Section \ref{sec:tele} we introduce
teleparallel gravity. In Section \ref{sec:lack} we explain the equivalence
between TEGR and GR, and the loss of local Lorentz invariance in $f(T)$
theories. We also analyze the survival of some TEGR solutions in $f(T)$
theories. In Section \ref{sec:null} we show that teleparallelism can be
formulated in terms of null tetrads; we exploit this fact to search for
surviving solutions. In Section \ref{sec:kerr} we show that Kerr geometry is
one of this kind of solutions. In Section \ref{sec:disc} we present the
conclusions.


\section{Teleparallel gravity}

\label{sec:tele}

Teleparallelism is a name for theories of gravity where the dynamical
variable is not the metric but the tetrad or \textit{vierbein}. The tetrad
field $\{\mathbf{e}_{a}(\mathbf{x})\}$ is a set of four orthonormal vectors
at each point $p$ of the manifold $M$ that constitutes a basis of the
tangent space $T_{p}M$. The dual co-frame $\{\mathbf{e}^{a}(\mathbf{x})\}$
is a basis of the co-tangent space $T_{p}^{\ast }M$. They can be decomposed
in a coordinate basis as
\begin{equation}
\mathbf{e}^{a}\ =\ e_{\mu }^{a}\,dx^{\mu }\quad \text{and}\quad \mathbf{e}%
_{a}\ =\ e_{a}^{\mu }\,\partial _{\mu }\ ,
\end{equation}%
where $e_{\mu }^{a}$ and $e_{a}^{\mu }$ are the respective components which
fulfill
\begin{equation}
e_{\mu }^{a}\,e_{b}^{\mu }\ =\ \delta _{b}^{a}\quad \text{and}\quad e_{\mu
}^{a}\,e_{a}^{\nu }\ =\ \delta _{\mu }^{\nu }\ .  \label{dual}
\end{equation}%
Greek indices $\mu ,\nu ,...=0,1,2,3$ indicate spacetime coordinates. Latin
indices $a,b,...=0,1,2,3$ are associated with the tangent space; we will
call them Lorentzian indices.

The orthonormality condition is the link between the tetrad and the metric
\begin{equation}
\eta _{ab}\ =\ g_{\mu \nu }\,e_{a}^{\mu }\,e_{b}^{\nu }\ ,\
\label{ortonormal}
\end{equation}%
where $\eta _{ab}=$diag$(1,-1,-1,-1)$. Equation (\ref{dual}) is used
for inverting this relation and getting the metric from the tetrad
\begin{equation}
g_{\mu \nu }\ =\ \eta _{ab}\,e_{\mu }^{a}\,e_{\nu }^{b}\;\;\;\;\text{or}%
\;\;\;\;g^{\mu \nu }\ =\ \eta ^{ab}\,e_{a}^{\mu }\,e_{b}^{\nu }\ ,
\label{metrica-g}
\end{equation}
then it is
\begin{equation}
\sqrt{-g}\ =\ \text{det}[e_{\mu }^{a}]\ =\ e\ .  \label{g-e}
\end{equation}%
Noticeably, the relation metric-tetrad is invariant under local Lorentz
transformations of the tetrad. In other words, there are many tetrad fields
for the same metric field. This also means that a dynamical theory for the
tetrad will determine the dynamic of the metric.

Teleparallelism is a dynamical theory for the tetrad whose Lagrangian is
built from the torsion tensor $T_{\ \ \nu \rho }^{\mu }$ associated with the
Weitzenb\"{o}ck connection \cite{Wei23}%
\begin{equation}
\overset{\text{{\tiny {W}}}}{\Gamma }\ _{\rho \nu }^{\mu }\ =\ e_{a}^{\mu }\
\partial _{\nu }e_{\rho }^{a}\ =\ -e_{\rho }^{a}\ \partial _{\nu }e_{a}^{\mu
}~,  \label{weitz}
\end{equation}%
\begin{equation}
T_{\ \ \nu \rho }^{\mu }\ =\ e_{a}^{\mu }\ (\partial _{\nu }e_{\rho
}^{a}-\partial _{\rho }e_{\nu }^{a})~.  \label{torsion}
\end{equation}%
Weitzenb\"{o}ck connection $\overset{\text{{\tiny {W}}}}{\Gamma }\ _{\rho
\nu }^{\mu }$ is curvatureless. Therefore, teleparallelism encodes gravity
in the torsion instead of the Riemann tensor. Weitzenb\"{o}ck connection has
the nice property that parallel-transported vectors keep constant their
projections on the vectors $\mathbf{e}_{a}$; in fact, it is $\overset{\text{%
{\tiny {W}}}}{\nabla }_{\nu }V^{\mu }=e_{a}^{\mu }\,\partial _{\nu
}(e_{\lambda }^{a}\,V^{\lambda })$. In particular, $\overset{\text{{\tiny {W}%
}}}{\nabla }_{\nu }e_{a}^{\mu }\equiv 0$, which means that Weitzenb\"{o}ck
connection is \emph{metric}. Since the curvature vanishes and the tetrad is
parallel-transported, one concludes that the tetrad field is a global frame.

In the teleparallel equivalent of GR, the tetrad is governed by the action
\cite{Hay79,Arc04,Mal13}
\begin{equation}
S_{TEGR}[\mathbf{e}_{a}]=\ \frac{1}{2\kappa }\ \int d^{4}x\,\ e\,\ S_{\rho
}^{\ \ \mu \nu }\,T_{\ \ \mu \nu }^{\rho }\ \ ,  \label{TEGR}
\end{equation}%
where $\kappa =8\pi G$ and $S_{\rho }^{\ \ \mu \nu }$ is defined as
\begin{equation}
2\,S_{\rho }^{\ \ \mu \nu }\equiv \underbrace{\frac{1}{2}\,(T_{\rho }^{\
\,\mu \nu }-T_{\ \ \,\,\rho }^{\mu \nu }+T_{\ \ \,\,\rho }^{\nu \mu })}_{%
\text{\textit{contorsion}}\ K_{\ \ \,\rho }^{\mu \nu }}\,+\,T_{\lambda }^{\
\,\lambda \mu }\ \delta _{\rho }^{\nu }\,-\,T_{\lambda }^{\ \,\lambda \nu }\
\delta _{\rho }^{\mu }\,.
\end{equation}%
The \emph{contorsion} $K_{\ \ \,\rho }^{\mu \nu }$ equals the difference
between Weitzenb\"{o}ck and Levi-Civita connections.

Analogously to $f(R)$ gravity \cite{Buc70, Sot10}, where the GR Lagrangian
is changed to an arbitrary function $f$ of the scalar curvature $R$, $f(T)$
theories constitute a teleparallel version of modified gravity obtained by
deforming the TEGR Lagrangian in terms of an arbitrary function of the \emph{%
Weitzenb\"{o}ck invariant} $T\equiv S_{\rho }^{\ \ \mu \nu }\,T_{\ \ \mu \nu
}^{\rho }$ \cite{Fer07,Fer08}. Its action reads
\begin{equation}
S[\mathbf{e}_{a}]\ =\frac{1}{2\kappa }\ \int d^{4}x\ \,e\,\ f(T)\ .
\end{equation}%
For minimally coupled matter, the dynamical equations of $f(T)$ gravity are
\begin{equation}
\ 4\ e^{-1}\ \partial _{\mu }(e\ \ e_{a}^{\lambda }\ S_{\lambda }^{\ \ \mu
\nu }\ f^{\prime }(T))\ +\ 4\ e_{a}^{\lambda }\ T_{\ \ \mu \lambda }^{\rho
}\ S_{\rho }^{\ \ \mu \nu }\ f^{\prime }(T)\ -\ e_{a}^{\nu }\ f(T)\ \ =\
-2\,\kappa \ e_{a}^{\lambda }\ \mathcal{T}_{\lambda }^{\ \nu }\ ,
\label{ecuaciones}
\end{equation}%
where $\mathcal{T}_{\lambda }^{\ \nu }$ is the energy-momentum tensor. As
seen, the dynamical equations are second order, which is a distinctive
feature regarding other theories of modified gravity. TEGR dynamics
corresponds to the particular case $f(T)=T$, $\ f^{\prime }(T)=1$.


\section{Comparing TEGR and $f(T)$ gravity}

\label{sec:lack}

The equivalence between GR and TEGR emanates from the following property:
Einstein-Hilbert Lagrangian differs from TEGR Lagrangian in a
four-divergence. In fact, by computing the Levi-Civita scalar curvature $R$\
for the metric (\ref{metrica-g}) one gets the result
\begin{equation}
-e\,\ R[\mathbf{e}_{a}]\ = \ e\,\ T-\ 2\ \partial _{\rho }(e\,T_{\ \ \mu
}^{\mu \ \ \rho })\ .  \label{equivalencia}
\end{equation}%
Therefore TEGR and Einstein-Hilbert actions are equivalent. This also means
that GR and TEGR harbor the same number of degrees of freedom. Even though
the tetrad field contains 16 components (6 more than the metric field), TEGR
is invariant under local Lorentz transformations of the tetrad. This gauge
freedom cancels out the excess of degrees of freedom. At the level of the
TEGR Lagrangian, a local Lorentz transformation is a local change to another
orthonormal basis,%
\begin{equation}
\mathbf{e}_{a^{\prime }}=\Lambda _{a^{\prime }}^{a}(\mathbf{x})\ \mathbf{e}%
_{a}\;\;\;\;,\;\;\;\;\mathbf{e}^{a^{\prime }}=\Lambda _{a}^{a^{\prime }}(%
\mathbf{x})\ \mathbf{e}^{a},
\end{equation}%
that adds a four-divergence to the TEGR Lagrangian $e\,T$. This behavior is
evident in Eq.~(\ref{equivalencia}) because $e\,R$ is invariant under local
Lorentz transformations.

Instead $f(T)$ gravity, like other theories of modified gravity, possesses
extra degrees of freedom. In fact, except for the case $f(T)=T$ (i.e., TEGR)
the dynamical equations (\ref{ecuaciones}) are sensitive to local Lorentz
transformations of the tetrad. This implies that the dynamical equations not
only contain information about the evolution of the metric but also about
some extra degrees of freedom exclusively associated with the tetrad that
are not present in the undeformed theory \cite{Yan10,Li11b,Li11,Li11a,Sot11}%
. At the level of the $f(T)$ Lagrangian, under a local Lorentz
transformation the Lagrangian changes as%
\begin{equation}
e\ \,f(T)\longrightarrow e\ \,f(T+\text{four-divergence})\ .
\end{equation}%
In this case the four-divergence term remains encapsulated inside
the function $f$ spoiling the local Lorentz invariance. The loss of
the local Lorentz invariance implies the existence of a preferential
global reference frame defined by the autoparallel curves of the
manifold that consistently solve the dynamical equations. That is,
Eqs.~(\ref{ecuaciones}) not only determine the metric but they also
choose some other characteristics of the tetrad field, so endowing
the spacetime with an absolute paralellization. The tetrads
connected by local Lorentz transformations lead to the same metric
but they are different with respect to the proper parallel
framework. Due to this essential feature of $f(T)$ theories, when
looking for solutions of a given symmetry it is quite complicated to
do an ansatz for the tetrad field. Actually, the symmetry helps us
to choose suitable coordinates to write the metric in a simple way.
But this does not say much about the ansatz for the tetrad. Indeed,
a very common mistake is to force $e_{\mu }^{a}$ to be diagonal in
the chosen coordinates. Frequently
this choice does not work as an ansatz for solving the equations (\ref%
{ecuaciones}); it is not consistent. For instance, in Ref.~\cite{Fer11a} it
was shown that a diagonal choice for FLRW universes only works in the flat
case; open and closed universes require non-trivial tetrads for solving the $%
f(T)$ dynamical equations.

In Ref.~\cite{Fer11b} it was proved that a naive diagonal tetrad does not
properly parallelize a static spherically symmetric geometry in $f(T)$
gravity. Therefore, we emphasize that the symmetries of the geometry are not
enough to visualize the absolute parallelization of the manifold, being
those quite futile in order to obtain the right answer. Certainly, in the
context of $f(T)$ theories, the proper frame which parallelizes the
spacetime for a given symmetry of the geometry must be independent of the
function $f$ \cite{Tam12}.

This article is aimed to find the parallelization for axially
symmetric solutions in $f(T)$ theories. In particular, we want to
know whether Kerr geometry survives or not in $f(T)$ gravity. To
answer the question we should find the correct ansatz to solve the
equations (\ref{ecuaciones}). This search is greatly facilitated by
invoking the following argument concerning the survival of certain
TEGR solutions \cite{Fer11b}: if a vacuum solution of $f(T)$ gravity
has $T=0$, then it will be a solution of TEGR as well (a
cosmological constant might be necessary). In fact, the replacement
of $T=0$ in Eq.~(\ref{ecuaciones}) leads to
\begin{equation}
4~e^{-1}\partial _{\mu }(e~e_{a}^{\lambda }~S_{\lambda }^{\,\,\mu \nu
})+4~e_{a}^{\lambda }~T_{\,\,\mu \lambda }^{\rho }~S_{\rho }^{\,\,\mu \nu
}-e_{a}^{\nu }\ \frac{f(0)}{f^{\prime }(0)}=0~,
\end{equation}%
which is a TEGR vacuum equation with cosmological constant $2\Lambda
=f(0)/f^{\prime }(0)$. We can avoid the cosmological constant term
by restricting the family of functions $f$ to those having $f(0)=0$,
$f^{\prime }(0)\neq 0$. In other words, we can exploit the freedom
to do local Lorentz transformations in TEGR to look for a tetrad
having $T=0$; if we success, then we will state that such solution
survives in $f(T)$ gravity. Notice that TEGR vacuum solutions does
not compels $T$ to vanish; $R$ must vanish. Thus
Eq.~(\ref{equivalencia}) says that $T$ is a four-divergence. So, the
former argument is based on the sensitivity of $T$ to local Lorentz
transformations. The above argument means that TEGR vacuum solutions having $%
T=0$ (or $T=$ constant) cannot be deformed by $f(T)$ gravity. We are going
to show that this is the case for Kerr geometry, what means that $f(T)$
gravity is unable to smooth the singularity of a black hole \cite{Fer11b}.

\section{Null tetrad approach}

\label{sec:null}

The search for a tetrad having $T=0$, if it exists, is easier by working
with a null tetrad. Any orthonormal tetrad $\{\mathbf{e}^{a}\}$ defines a
null tetrad $\{\mathbf{n}^{a}\}=\{\mathbf{l},\mathbf{n},\mathbf{m},\overline{%
\mathbf{m}}\}$
\begin{equation}
\mathbf{l}=\frac{(\mathbf{e}^{0}+\mathbf{e}^{1})}{\sqrt{2}},\quad \mathbf{n}=%
\frac{(\mathbf{e}^{0}-\mathbf{e}^{1})}{\sqrt{2}},\quad \mathbf{m}=\frac{(%
\mathbf{e}^{2}+i\,\mathbf{e}^{3})}{\sqrt{2}},\quad \overline{\mathbf{m}}=%
\frac{(\mathbf{e}^{2}-i\,\mathbf{e}^{3})}{\sqrt{2}}\ .
\label{complex-transf}
\end{equation}%
This tetrad form a null basis
\begin{equation}
\mathbf{l}\cdot \mathbf{l}=0,\;\;\mathbf{n}\cdot \mathbf{n}=0,\;\;\mathbf{m}%
\cdot \mathbf{m}=0,\;\;\overline{\mathbf{m}}\cdot \overline{\mathbf{m}}=0\ ,
\label{null1}
\end{equation}%
but it is not orthornormal
\begin{equation}
\mathbf{l}\cdot \mathbf{n}=1,\;\;\mathbf{m}\cdot \overline{\mathbf{m}}%
=-1,\;\;\mathbf{l}\cdot \mathbf{m}=0,\;\;\mathbf{n}\cdot \mathbf{m}=0\ .
\label{null2}
\end{equation}%
We can solve $\{\mathbf{e}^{a}\}$ in Eq.~(\ref{complex-transf}) and replace
in Eq.~(\ref{metrica-g}) to get the metric in terms of a null tetrad
\begin{equation}
g_{\mu \nu }\ =\ \eta _{ab}\,n_{\mu }^{a}\,n_{\nu }^{b}~,
\end{equation}%
where $\eta _{ab}$ is now
\begin{equation}
\eta _{ab}=\left(
\begin{array}{cccc}
0 & 1 & 0 & 0 \\
1 & 0 & 0 & 0 \\
0 & 0 & 0 & -1 \\
0 & 0 & -1 & 0%
\end{array}%
\right) \ .
\end{equation}%
Therefore, the metric reads
\begin{equation}
\mathbf{g}=\mathbf{n}\otimes \mathbf{l}+\mathbf{l}\otimes \mathbf{n}-\mathbf{%
m}\otimes \overline{\mathbf{m}}-\overline{\mathbf{m}}\otimes \mathbf{m}\ .
\label{null-metric}
\end{equation}
We are going to transform the tetrad $\{\mathbf{e}^{a}\}$ of a given
TEGR vacuum solution to look for a tetrad having $T=0$. This
procedure is equivalent to change $\{\mathbf{n}^{a}\}$. Since the geometry (\ref%
{null-metric}) and the relations (\ref{null1}), (\ref{null2}) cannot be
changed, a simple try is
\begin{equation}
\mathbf{l}\longrightarrow \ \exp [\lambda (\mathbf{x})]\ \mathbf{l}\ ,\ \ \
\ \ \ \ \ \ \ \ \mathbf{n}\longrightarrow \ \exp [-\lambda (\mathbf{x})]\
\mathbf{n}\ .  \label{null-transf}
\end{equation}%
This change implies a local Lorentz boost along the direction of
$\mathbf{e}^{1}$ with parameter $\gamma (\mathbf{x})=\cosh [\lambda
(\mathbf{x})]$.

Since $R=0$ for vacuum solutions, and we are looking for solutions having $%
T=0$, then Eq.~(\ref{equivalencia}) states that the four-divergence will
vanish as well%
\begin{equation}
\partial _{\rho }(e\,T_{\ \ \mu }^{\mu \ \ \rho })=0~.
\end{equation}%
Remarkably, Weitzenb\"{o}ck torsion (\ref{torsion}) does not change under
global linear transformations of the basis. This implies that even the null
tetrad can be used to compute $T_{\ \ \nu \rho }^{\mu }$%
\begin{equation}
T_{\ \ \nu \rho }^{\mu }\ =\ n_{a}^{\mu }\ (\partial _{\nu }n_{\rho
}^{a}-\partial _{\rho }n_{\nu }^{a})~.
\end{equation}%
Under the transformation (\ref{null-transf}), the vector sector of torsion
(the one appearing in the four-divergence) changes as%
\begin{equation}
T_{\ \ \mu }^{\mu \ \ \rho }\longrightarrow T_{\ \ \mu }^{\mu \ \ \rho
}+(l^{\mu }l^{\rho }-n^{\mu }n^{\rho })~\partial _{\mu }\lambda (\mathbf{x}%
)~.
\end{equation}


\section{Kerr geometry with vanishing $T$}

\label{sec:kerr}

In a proper chart, Kerr geometry \cite{Ker65,Debney67} reads
\begin{eqnarray}
ds^{2} &=&\left( 1-\dfrac{2\,m\,r}{\Sigma }\right) \ dt^{2}+\dfrac{4\,m\,r}{%
\Sigma }\ dt\,dr+\dfrac{4\,a\,m\,r\ \sin ^{2}\theta }{\Sigma }\ dt\,d\varphi
-\left( 1+\dfrac{2\,m\,r}{\Sigma }\right) \ dr^{2}  \notag \\
&&-2\,a\ \sin ^{2}\theta \ \left( 1+\dfrac{2\,m\,r}{\Sigma }\right) \
dr\,d\varphi -\Sigma \ d\theta ^{2}-\left[ \Sigma \ \sin ^{2}\theta +\left(
1+\dfrac{2\,m\,r}{\Sigma }\right) a^{2}\sin ^{4}\theta \right] \ d\varphi
^{2}\ ,  \label{kerr}
\end{eqnarray}%
where $m$ is the mass of the black hole and $\Sigma =r^{2}+a^{2}\cos
^{2}\theta $, $a$ being the angular momentum per unit of mass. In this
expression, coordinates $x^{\mu }=(t,r,\theta ,\phi )$ are linked to the
usual Boyer-Lindquist coordinates $\tilde{x}^{\mu }=(\tilde{t},r,\theta ,%
\tilde{\phi})$ through the relations
\begin{equation}
d\tilde{t}=dt+\frac{2\,m\,r}{r^{2}+a^{2}-2\,m\,r}\,dr\quad \text{and}\quad d%
\tilde{\phi}=d\phi +\frac{a}{r^{2}+a^{2}-2\,m\,r}\,dr~.  \label{varphi-time}
\end{equation}%
The geometry (\ref{kerr}) can be written in the way
(\ref{null-metric}) by using the null tetrad
\begin{equation}
\mathbf{n}_{\ \,\mu }^{a}=\dfrac{1}{\sqrt{2}}\left(
\begin{array}{cccc}
\exp [\lambda ] \left( 1-\dfrac{2\,m\,r}{\Sigma }\right)  & \exp
[\lambda
] \left( 1+\dfrac{2\,m\,r}{\Sigma }\right)  & 0 & \exp [\lambda ] \left( 1+%
\dfrac{2\,m\,r}{\Sigma }\right) \,a\,\sin ^{2}\theta  \\
\exp [-\lambda ]  & -\exp [-\lambda ]  & 0 & -\exp [-\lambda ]\
a\,\sin
^{2}\theta  \\
0 & 0 & r+i\,a\cos \theta  & (r+ia\cos \theta )\,i\,\sin \theta  \\
0 & 0 & r-i\,a\cos \theta  & -(r-ia\cos \theta )\,i\,\sin \theta
\end{array}%
\right) \ ,
\end{equation}%
where $\lambda =\lambda (t,r,\theta )$. Then, the Weitzenb\"{o}ck invariant
becomes
\begin{equation}
T=\frac{2}{\Sigma ^{3}}\ \left( \Sigma ^{2}-\ 4\,a^{2}\cos
^{2}\theta \ (\Sigma +\,m\,r)-\ 2\,r\ \Sigma ^{2}\ \partial
_{t}\lambda \right) \ .\label{T}
\end{equation}%
So, there is a family of functions $\lambda (t,r,\theta )$ that realize the
vanishing of $T$
\begin{equation}
\lambda (t,r,\theta )\ =\ \frac{t}{2\,r}\ \left( 1-\ 4\ a^{2}\cos
^{2}\theta \ \frac{\Sigma +\,m\,r}{\Sigma ^{2}}\right) +\lambda
_{1}(r,\theta )\ .\label{lambdaK}
\end{equation}%
Therefore, Kerr geometry is a solution of $f(T)$ gravity. In
Ref.~\cite{Nas14}, an axially symmetric tetrad having null $T$ was
also found by directly solving the vacuum $f(T)$ equations. That
Kerr tetrad not necessarily coincides with the one here obtained due
to remnant symmetries characterizing $f(T)$ solutions \cite{Fer15}.

Notice that the freedom to choose the function $\lambda_1$ can be
exploited for replacing $t$ with $\tilde{t}$ in Eqs.~(\ref{lambdaK})
and (\ref{lambdaS}). It should be also pointed out that the function
$\lambda$ is not well defined at $r=0$. In Kerr geometry, the region
$r=0$ is a circle where $\theta$ plays the role of radial
coordinate. The edge of the circle ($r=0$ and $\theta = \pi/2$) is
the Kerr ring singularity, but its inner region is not singular. The
solution in this region should be re-elaborated according to the
radial meaning of $\theta$ coordinate. Of course, Schwarzschild
tetrad is free of this problem because the singularity at $r=0$ is
just a point. In this case, the function  $\lambda $ becomes
\begin{equation}
\lambda (t,r)=\frac{t}{2\ r}+\lambda _{1}(r)\ .\label{lambdaS}
\end{equation}


\section{Final comments}

\label{sec:disc}

We have proved that Kerr geometry survives as a solution of $f(T)$ gravity
whenever the function $f$ satisfies $f(0)=0$ and $f^{\prime }(0)\neq 0$. We
invoked the argument that any GR vacuum solution will remain a solution of $%
f(T)$ gravity if it admits a tetrad for which the Weitzenb\"{o}ck invariant $%
T$ vanishes. This argument was used in Ref.~\cite{Fer11b} for
showing that Schwarzschild geometry survives in $f(T)$ theories.
Here we showed that the use of null tetrads can help to easily prove
the existence of a tetrad with vanishing $T$ even if the symmetry is
not spherical but axial. We remark that $T$ remains null when
passing from the null tetrad to an orthonormal tetrad by means of
the relations (\ref{complex-transf}). This is because torsion $T_{\
\ \nu \rho }^{\mu }$ is invariant under global linear
transformations of the basis.

\bigskip

It should be emphasized that the simplicity of the demonstration
given in Section \ref{sec:kerr} relies on a good choice of the null
tetrad. We started from the null tetrad associated with the
Kerr-Schild form of Kerr metric \cite{Ker65,Debney67}
\begin{equation}
\mathbf{g}\ =\ \mathbf{n}_o\otimes (\mathbf{l}_o + f\
\mathbf{n}_o)+(\mathbf{l}_o
+ f\ \mathbf{n}_o)\otimes \mathbf{n}_o-\mathbf{%
m}_o\otimes \overline{\mathbf{m}}_o-\overline{\mathbf{m}}_o\otimes
\mathbf{m}_o\ =\ \mathbf{g}_o+2\, f\ \mathbf{n}_o\otimes
\mathbf{n}_o \label{KS}
\end{equation}
(i.e., $\mathbf{l}=\mathbf{l}_o + f\ \mathbf{n}_o$), where
$\{\mathbf{n}^{a}_o\}$ is a suitable null tetrad for Minkowski
metric $\mathbf{g}_o$ that leaves Kerr gravity completely encoded in
the function $f$
\begin{equation}
f(r,\theta)\ =\ -\frac{2\, m\, r}{\Sigma}\ .
\end{equation}
Then we applied the ansatz (\ref{null-transf}), and obtained a
differential equation for $\lambda$ by demanding the vanishing of
the Weitzenb\"{o}ck invariant (\ref{T}). This differential equation was
particularly simple because of a good choice of coordinates: we used
the chart employed in the Newman-Janis algorithm for passing from
Schwarzschild solution to Kerr solution \cite{Fer14}.

\bigskip

The obtained results show that $f(T)$ theories are unable to smooth
black hole singularities. It would be of major interest to know
whether this conclusion can be extended to more general geometries.
On the other hand, it is known that this kind of singularities can
be smoothed by means of a different scheme of modified teleparallel
gravity \cite{Fer11a}.


\begin{acknowledgments}
This work was supported by Consejo Nacional de Investigaciones Cient\'{\i}%
ficas y T\'{e}cnicas (CONICET) and Universidad de Buenos Aires. The authors
thank Franco Fiorini for helpful comments.
\end{acknowledgments}


\end{document}